\documentclass{aa}  
\usepackage{natbib} 
\bibpunct{(}{)}{;}{a}{}{,} 
\usepackage{graphicx}
\begin{document}
   \title{Temporal relation between quiet-Sun transverse fields and the strong flows detected by IMaX/SUNRISE}


   \author{C. Quintero Noda
          \inst{1,2}
	  \and
          V. Mart\'inez Pillet\inst{1,2}
          \and
          J.M. Borrero\inst{3}	  
          \and
	  S. K. Solanki\inst{4,5}
          }

   \institute{Instituto de Astrof\'isica de Canarias, E-38200, La Laguna, Tenerife, Spain.\quad \email{cqn@iac.es}
	\and
         Departamento de Astrof\'isica, Univ. de La Laguna, La Laguna, Tenerife, E-38205, Spain
	\and
	 Kiepenheuer-Institut f\"ur Sonnenphysik, Sch\"oneckstr. 6, D-79104, Freiburg, Germany
         \and
	Max-Planck-Institut f\"ur Sonnensystemforschung, Max-Planck-Str. 2, 37191 Katlenburg-Lindau, Germany
     \and
     School of Space Research, Kyung Hee University, Yongin, Gyeonggi 446-701, Republic of Korea}

   \date{Received , 2013; accepted , }

 
  \abstract
   {Localized strongly Doppler-shifted Stokes $V$ signals were detected by IMaX/SUNRISE. These signals are related to newly emerged magnetic loops that are observed as linear polarization features.}
   {We aim to set constraints on the physical nature and causes of these highly Doppler-shifted signals. In particular, the
temporal relation between the appearance of transverse fields and the strong Doppler shifts is analyzed in some detail.}
   {We calculated the time difference between the appearance of the strong flows and the linear polarization. 
We also obtained the distances from the center of various features to the nearest neutral lines and
whether they overlap or not.
These distances were compared with those obtained from randomly distributed points on observed magnetograms.  
Various cases of strong flows are described in some detail.}
   {The linear polarization signals precede the appearance of the strong flows by on average 84$\pm$11 seconds.
The strongly Doppler-shifted signals are closer (0.\arcsec19) to magnetic neutral lines than 
randomly distributed points (0.\arcsec5). Eighty percent of the strongly Doppler-shifted signals are close to a
 neutral line that is located between the emerging field and pre-existing fields.
That the remaining 20\% do not show a close-by pre-existing field could be explained by a lack of sensitivity or an unfavorable geometry of the pre-existing field, for instance, a canopy-like structure.} 
   {Transverse fields occurred before the observation of the strong Doppler shifts. The process is most naturally
explained as the emergence of a granular-scale loop that first gives rise to the linear polarization signals,
interacts with pre-existing fields (generating new neutral line configurations), and produces the observed strong
flows. This explanation is indicative of frequent small-scale reconnection events in the quiet Sun.}
\titlerunning{The temporal relation between transverse fields and quiet Sun jets.}          
                             
                             
\authorrunning{Quintero Noda et al.}

   \maketitle
   


\section{Introduction}

The analysis of the quiet-Sun magnetism has been enormously advanced in the
past years \citep[see, e.g.,][for recent reviews]{dewijn2009, martinezpillet2013}. 
A crucial aspect in this achievement has
been the successful performance of instruments onboard the Hinode 
\citep[see][]{Kosugi2007,Tsuneta2008} and SUNRISE missions
\citep[see][]{Barthol2011,Solanki2010}. They have both provided quantitative
data with spatial resolutions similar to or better than those available in the past and
very significantly improved the temporal consistency, going beyond the typical granular evolutionary timescales.
Supplemented with a polarimetric sensitivity of, typically, one part in $10^3$,
these observational conditions have allowed the characterization of the horizontal
component of quiet-Sun fields. It has been found that they typically appear at
locations with intermediate continuum intensities
\citep[]{Lites2008,Danilovic2010_2}, near the edges of granules \citep[as
originally found by the Advanced Stokes Polarimeter, ASP][]{Lites1996}, and
have a clear transient nature \citep{Ishikawa2009,Danilovic2010_2}.  The temporal evolution of
these high-quality polarimetric data can often be described as loops emerging
at the surface 
\citep{MartinezGonzalez2007,Centeno2007}, protruding into various heights in the atmosphere
\citep{MartinezGonzalez2009}, and with a characteristic non-uniform spatial
distribution of the emerging loops that displays `dead calm' areas \citep[][who
previously reported on these non-uniform
distributions]{MartinezGonzalez2012, Lites2008}.

High-speed Doppler-shifted signals have been found in a number of cases in the quiet-Sun fields,
mostly seen in the Stokes $V$ parameter \citep[see a variety of examples in][]{Shimizu2008}.
The convective-collapse process
first proposed on theoretical grounds by \citet{Parker1978}, \citet{Webb1978}, and \citet{Spruit1979} 
has been tentatively identified in the works of
\citet{BellotRubio2001} and \citet{Nagata2008} as the amplification of the
magnetic field induced by a strong downflow of cold material \citep{Danilovic2010_1}.
The line-of-sight (LOS) component of these downflows reaches speeds of 6 km s$^{-1}$.
It has also been proposed that these downflows hit dense layers and
rebound into upflows with supersonic speeds as well \citep{Grossmann1998}, which would explain the extremely
blueshifted signals observed by \citet{SocasNavarro2005}.

Perhaps the most unexpected discovery made with the IMaX/SUNRISE instrument
\citep[see][]{MartinezPillet2011b} is the detection of subarcsecond patches 
of circular polarization at the continuum reference wavelength point of this 
instrument \citep{Borrero2010}. In its normal observing mode, this magnetograph 
maps the Fe~{\sc i} line in 5250.2 \AA~in four wavelength points within the line and one in the
continuum region mid-range between this line and the neighboring Fe~{\sc i} 5250.6 \AA~line.
The exact continuum point is at 227 m\AA~from the IMaX line center. As stated by 
\citet{Borrero2010}, the only way to generate circular polarization signals at this
continuum wavelength is by Doppler-shifting either one of the Fe~{\sc i} line signals there. 
The fact that most of the observed continuum polarization patches occurred on top of
upflowing granules led these authors to suggest that the patches were produced by upflows that shift the
Fe~{\sc i} 5250.6 \AA~line to the continuum point. The estimated LOS velocities needed to generate 
these shifts were in the range [5,12] km s$^{-1}$\citep{Borrero2012}. Because both opposite polarity patches and
transverse fields were seen close to the location of the continuum polarization 
patches, some form of reconnection between newly emerged loops and surrounding fields was 
proposed as the most plausible scenario for their occurrence. More recently, \citet{Borrero2013} 
have used data from the same instrument, but with better spectral coverage, to perform
inversions that result in increased heating in the upper layers and in opposite polarities along the LOS. 
This result further consolidates the hypothesis of a 
reconnection-driven process. The presence of these highly Doppler-shifted signals 
at what, otherwise, are continuum wavelengths has been 
confirmed in Hinode/SP data, as shown by \citet{MartinezPillet2011a}. The
superior spectral sampling of the Hinode/SP instrument allowed a clear separation of
blueshifted and redshifted events (which these authors termed quiet-Sun jets), which
was not possible with the single-continuum point of IMaX/SUNRISE. These authors
confirmed the transverse field regions in the surroundings of the jets and
found a tendency for quiet-Sun jets to occur in blue- and redshifted pairs.

The chronological relation between the transverse field patches and the jets 
remains unclear from these studies
\citep[we continue the naming convention proposed by][]{MartinezPillet2011a}:
Is there a preference for the transverse signals or for the jets to occur first?
If reconnection is driven by the emergence of new magnetic bipoles into the solar
atmosphere, it is expected that the transverse signals {\em precede} the detection
of the jets. Taking advantage of the large FOV and good temporal cadence
of the IMaX/SUNRISE data, we study
in this work the temporal sequence of the detection of the transverse fields
and the jets. Other aspects of this phenomenon, such as the distance of the
jets to the nearest neutral line and some case examples of transverse fields that 
preceded the jets, are discussed at some length.

\begin{figure}
\centering
\includegraphics[width=8.8cm]{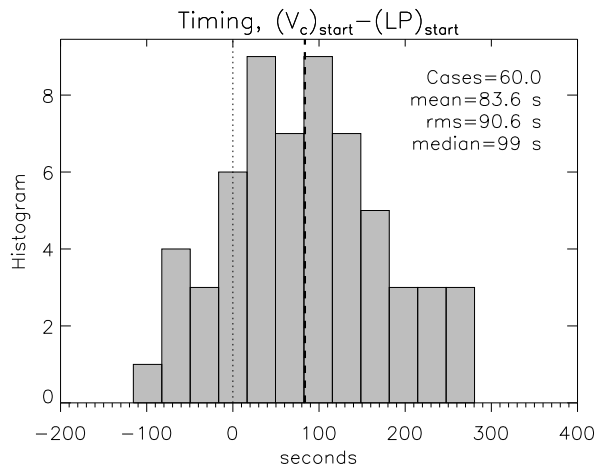}
\caption{Histogram of the time difference between the 
beginning of the $V_c$ jet signal and the
appearance of the linear polarization signal. Positive values mean that the linear
polarization precedes the appearance of the highly Doppler shifted signal. The thick dashed line represents 
the averaged time difference.}
\label{fig1}
\end{figure}
  
\section{Observations and data analysis}

We used the same two data sets as \citet{Borrero2010}. They were 
recorded on June 9, 2009 and have a duration of
22.6 and 31.9 minutes, respectively. The field of view (FOV) is
$47^{\prime\prime}\times 47^{\prime\prime}$. The data were recorded in the IMaX
V5-6 mode \citep{MartinezPillet2011b}, which sampled five wavelength positions
relative to $\lambda_{0}$ = 5250.217 \AA : $-80, -40, 40, 80$ m\AA, and a
fifth wavelength placed at the continuum: $\lambda_c=\lambda_0+227$ m\AA. \ The
instrument obtained the four Stokes parameters $(I,Q,U,V)^{\dag}$ with a time
cadence of 33 seconds and a pixel size of 0.$^{\prime\prime}055$. Six accumulations
were summed in this mode to achieve a signal-to-noise ratio of 10$^3$. The instrument
made periodic calibrations of the optical aberrations by using a phase
diversity glass plate that generates a known out-of-focus configuration
in one of the two cameras used to measure the orthogonal polarization states. 
These calibrations can be used to perform an image restoration of the data and bring
it closer to the diffraction limit. This restoration process also increases
the noise, so that for some detection purposes it may not be as well suited as the
non-restored data. Specifically, the noise increases by a factor 3 in the restored frames.  
Indeed, to identify the nearly horizontal fields, we used the non-restored linear polarization
signal averaged over the four wavelength positions in the spectral line, as in \citet{Danilovic2010_2}. 
The above-mentioned increase of noise in the restored $Q$ and $U$ frames allows only the stronger
linear polarization patches to be detected in the restored data.
The strong flows were detected also using the
non-restored circular polarization signal at $\lambda_c$. However, when using 
the normal magnetograms to identify the nearest neutral-line both, the non-restored
and restored data were included in the analysis. 

The events (jets and transverse fields) were detected manually. 
We established their intensity and size thresholds to identify them.
 First, we include in the analysis
linear polarization signals above 0.26$\%$ of the continuum level and circular
polarization signals at the continuum wavelength, $V_c$, above 0.4$\%$
(also normalized to the continuum intensity). 
Second, the selected events had
to reach a minimum area of five pixels in $V_c$ and the corresponding linear polarization 
patch sometime during their evolution. 
We used the identification of a jet event in $V_c$ as reference. Then 
we searched for a nearby, concurrent linear
polarization patch. This patch could occur before or after the first detection of the
highly Doppler-shifted signal.

The total number of jets found is 96, 72 of which (75$\%$) show a clear
relation with a linear polarization patch, while the rest of the events does not
show a connection with any enhanced linear polarization region. These results 
closely agree with the work of \citet{Borrero2010}, who used the restored data
to characterize properties such as the physical 
dimensions of the events for which the nearly 
diffraction-limited data are better suited.

\begin{figure*}
\centering
\includegraphics[width=18cm]{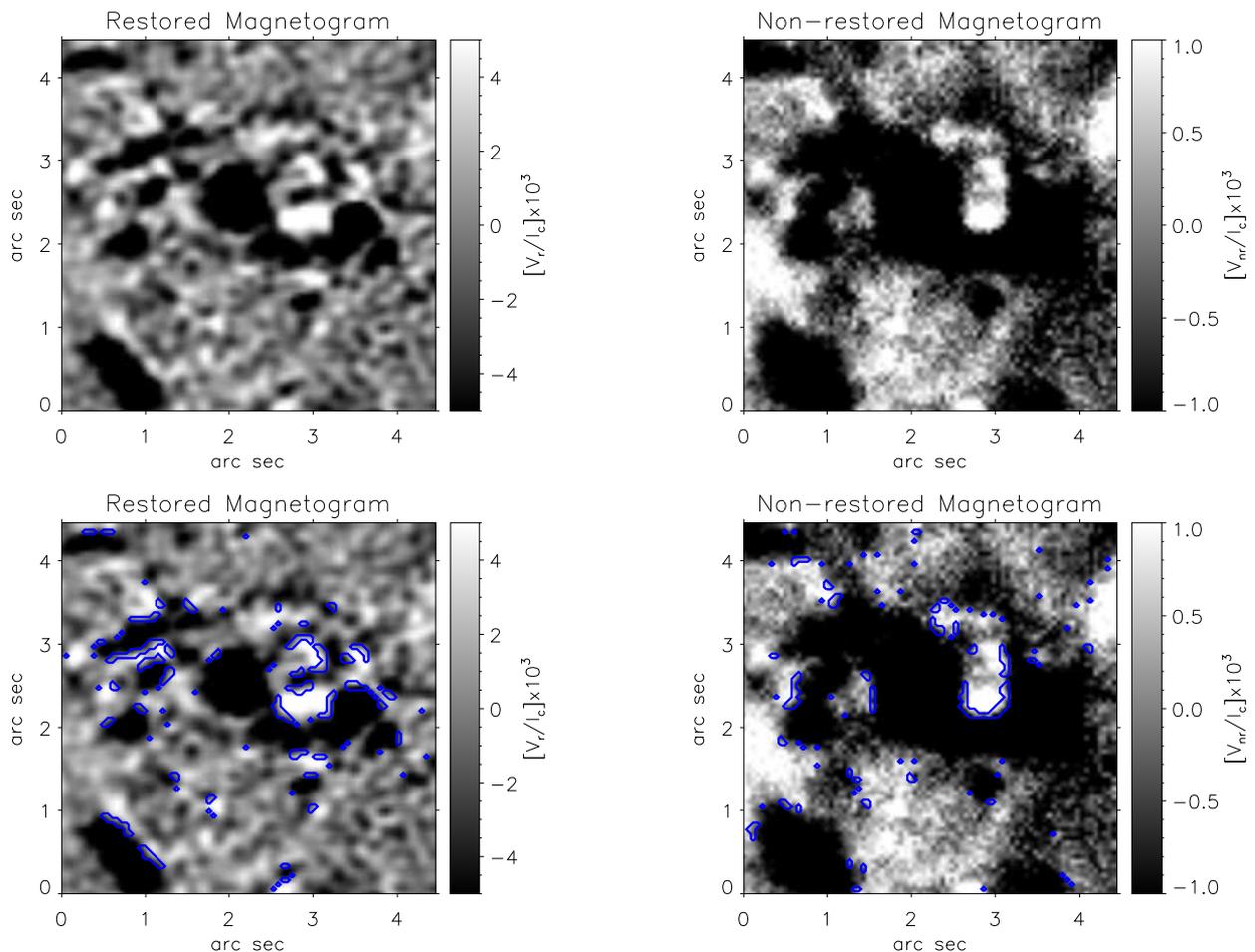}
\caption{
Example of neutral line identification in one IMaX/SUNRISE magnetogram.
The left column corresponds to restored and the right column to non-restored
magnetograms (but they represent the same observations). The top row displays the original magnetograms, 
the bottom row overlays the neutral lines (blue contours) identified by 
the edge-enhancement procedure explained in the text. Note the increased
frequency of neutral lines occurrences  in the restored data.
}
\label{figneutral}
\end{figure*}

\begin{figure*}
\centering
\includegraphics[width=18cm]{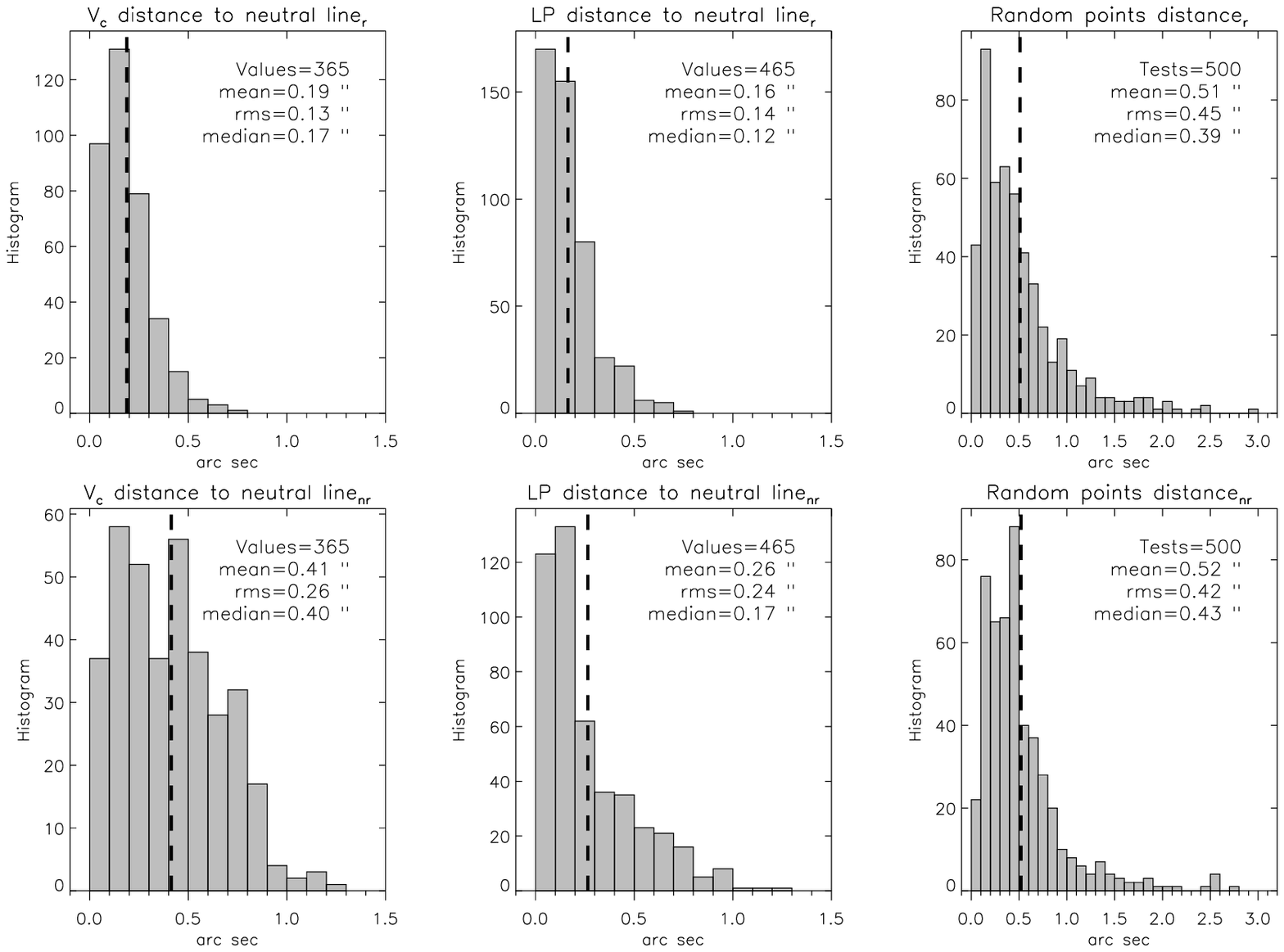}
\caption{Histograms of distances between various features and neutral lines.
The first column shows the distances between the centers of the
non-restored $V_c$ jets to the neutral lines defined on the restored (top) and on the non-restored magnetogram (bottom). 
The second column shows the same quantity for the linear polarization
patches. The third column presents the distances
of randomly distributed points to their nearest neutral lines. 
}
\label{fig2}
\end{figure*}

\section{Results}

The aim of this section is to study the temporal relation between the jets and
the associated linear polarization patches. We also quantify the typical
distances between these features and their nearest neutral line. Because
in the quiet-Sun observed by IMaX/SUNRISE neutral lines are frequently 
encountered, a comparison of these distances with those found for
randomly defined locations is mandatory.  These studies can help us to stablish a possible
scenario for the occurrence of the quiet-Sun jets and complement
the works of \citet{Borrero2010,Borrero2012} and \citet{MartinezPillet2011a}.

\subsection{Temporal relation between $V_c$ jets and the associated linear polarization regions}

We first quantified the time difference between the initial instant at which we
saw the signal at $V_c$ and the first moment at which we detected the
associated transverse field patch. Some of the 72 cases of quiet-Sun jets
associated to linear polarization signals were not included in the analysis
because the associated transverse field region was already present at the beginning 
of the time series. For these cases only a lower limit could be estimated and
we decided not to include them.
A precise time interval could be defined for only 60 cases.  

The histograms of the time differences in Fig. \ref{fig1} reveal that the associated linear
polarization patches precede the strong $V_c$ jets in most of the cases, 
by 84$\pm$11 seconds on average. It seems clear therefore that the Doppler-shifted signals are
a consequence of the small-scale flux emergence processes, as shown by the linear
polarization patches. 
Only 13$\%$ of the cases show a negative value, indicating that the jets
are seen before the corresponding transverse field region. It is important, however, to
bear in mind that the two types of signals are highly affected by visibility
effects. On the one hand, the $V_c$ signals are enhanced when a large component
of the underlying flow is aligned with the LOS. On the other hand, the linear polarization signals
are always very sensitive to the exact field line configuration and strength.
As a consequence, jets associated with small-scale flux emergence could be more
frequent than determined here \citep[or in][]{Borrero2010} and, often, we might not detect the 
linear polarization signals simply because of low field strengths, small features, or
both, resulting in low flux entities that are hidden 
below the noise level. These visibility problems might help explain the relatively
large standard deviation observed in Fig. \ref{fig1} (91 seconds). We also calculated 
the median of the data sets because the
shape of the distribution is not always clearly symmetric. We provide the opportunity to
 examine the results from the mean and the median values
of the distributions, although these two values are very close throughout the analysis.

\subsection{Distance to the nearest neutral line}

The fact that the jets have a clear preference to occur {\em after} the
emergence of the small-scale loop indicates that the evolution
of the loop once it is at the surface and its interaction with possible surrounding
fields might trigger the process. The existence of these granular scale
magnetic loops in the photosphere has already been clearly established by
various observations \citep[]{DePontieu2002, Centeno2007,MartinezGonzalez2009, 
Ishikawa2009b,Ishikawa2010}. In the previous studies of
\citet{Borrero2010,Borrero2012} and \citet{MartinezPillet2011a} all evidence
pointed to the possibility that some form of reconnection of the newly emerged loops
with pre-existing fields triggers the high-speed flows.
To additionally consolidate this possibility, we study in this section the
distance of the $V_c$ events to the nearest neutral line in the FOV. The location of
a jet event is defined by an intensity-weighted position estimated by  

\begin{equation}
\vec{r_c}=\frac{\sum_i \vec{r_i}\cdot |V_c|}{\sum_i |V_c|} \ ,
\label{eq1}
\end{equation}
where $r_i$ is the distance from each pixel of the event that displays a $V_c$ signal to
a reference system's origin defined in a small FOV of $100\times100$ pixels, or 
$5^{\prime\prime}.5\times5^{\prime\prime}.5$, which was used to individually analyze
the events.
From this location we then perform an automated search for the closest neutral line 
and measured the smallest distance between them. Neutral lines are identified in
edge-enhanced magnetograms that can be easily thresholded to display the pixels
where neutral lines are found. For this purpose we used the IDL procedure 
\textit{sobel.pro}, which is based on the Sobel edge-enhancement operator. It is a local 
non-linear operator that detects gradients between pixels in an image to give information
 about the edges of the structures displayed in that image. 
 An example of the selected pixels corresponding to our definition of the neutral line configuration
is presented in Figure \ref{figneutral}. 
Pixels inside the blue contours are considered as neutral lines. 
The distance from the locations of $V_c$ events (as defined
in Eq. \ref{eq1}) to the nearest neutral line is given by the closest pixel inside
the blue line contours encountered in the neighborhood.

The restored magnetograms offer a much higher resolution for the detection of neutral lines
than the non-restored ones, and the former often show neutral lines that are not seen in the latter.
But because we define the locations of $V_c$ jets and linear polarization patches
only in the non-restored magnetograms, we used 
the neutral lines found in the non-restored magnetograms in addition to those from
the restored magnetograms.
The center of the transverse patches was defined in a
similar way as for the $V_c$ signals defined in Eq. \ref{eq1}, but now weighting with
the amount of the observed linear polarization. 
We also estimated distances of randomly distributed points over the FOV to their nearest neutral lines
to compare with the obtained distances of the jets and of the linear polarization patches to the nearest neutral line.
Comparing the distances measured for linear polarization patches
and $V_c$ jets is interesting because for the first ones 
(which represent magnetic loops that always contain a
neutral line) we expect to deduce smaller distances than for the $V_c$
events. Similarly, randomly distributed points are expected to provide distances to
the nearest neutral line that gauge how likely it is to be close to one such
line in the IMaX magnetograms by chance. If the jet events are related to magnetic
reconnection processes, we expect them to be closer to the neutral lines than
the randomly distributed points. Note that this step is not needed to evaluate
the possible effect of erroneous detections of $V_c$ jets and/or neutral lines, but 
as a way of establishing a statistically meaningful reference of a spatial correlation 
between these features.

Jet events are often detected in several IMaX snapshots, for each of them
an estimate of the distance to the nearest neutral line is obtained. They
evolve in time in a way that shows proper motions \citep[see][]{Borrero2010}
that reflect the evolution of the underlying granulation, it is not clear which 
instant is more representative of the proximity to a neutral line. Thus, all of the 
above estimates were included in the analysis.
For these reasons, the 60 associated cases produce a total of 365 distance
values for the $V_c$ jets and
465 values for the associated linear polarization patches. Accordingly, we 
chose to study 500 randomly placed points to provide a similar statistical comparison.
 
Figure \ref{fig2} provides the histograms of the
various distances obtained from the non-restored  $V_c$ jets,
the non-restored linear polarization patches, and the randomly placed points
to the neutral lines detected on the restored and 
non-restored magnetograms (top and bottom rows, respectively). 
The histograms for the randomly
distributed points show that the typical distance to a neutral line in an
IMaX/SUNRISE magnetogram is only 0.\arcsec5 (in both the restored and
non-restored magnetograms).  However, the typical distance
of a $V_c$ jet is much smaller, around 0.\arcsec19 for the reconstructed
data. In agreement with
the predictions above, the average distance of the center of the linear
polarization patches to the nearest neutral line is smaller, about 0.\arcsec16.
This number nicely coincides with the spatial resolution of the data
\citep{MartinezPillet2011b}. Interestingly, the same ordering in the magnitudes
of the distances is obtained for the non-restored data, with the random points
located farthest from the neutral lines, followed by the jets and finally the linear
polarization signals as the closest ones. This consolidates the same trend as was
found for the restored magnetograms and demonstrates that the $V_c$ jets
detected by IMaX/SUNRISE are significantly closer to quiet-Sun neutral lines
than randomly selected points. Thus, these results additionally
support the hypothesis that these jets originate from reconnection 
events in the photosphere that are associated with, and usually follow,
the emergence of small-scale quiet-Sun loops.

\subsection{Relative distances between $V_c$ jets and transverse field patches}

\begin{figure}
\centering
\includegraphics[width=8.8cm]{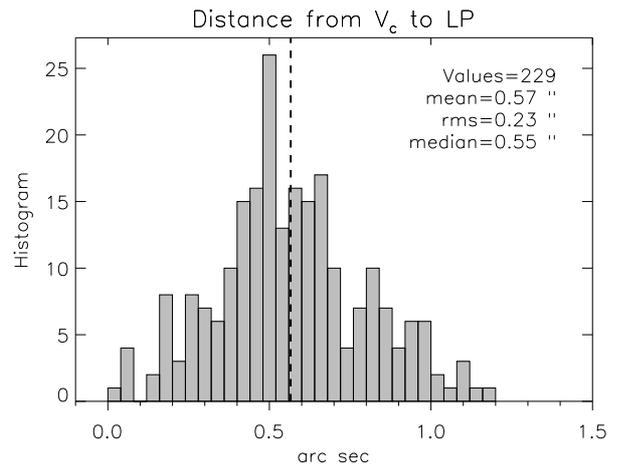}
\caption{Distances between $V_c$ jets and the associated linear polarization patches
as measured in the non-restored data.}
\label{fig3}
\end{figure}

With the above estimates, we studied the distance distribution of
the quiet-Sun jets and the associated linear polarization patches. Note
that in \cite{Borrero2010} this distance was not estimated and the
association between horizontal patches and $V_c$ events was considered real as
long as it was smaller than 2\arcsec. Of all locations measured in the
different snapshots, a total of 229 snapshots offered a clear correspondence
between jets and transverse fields, allowing an estimate of their separation.
The total number of distances is lower than the number used in Figure
\ref{fig2}. The reason is that almost every associated event presents a linear
polarization patch that occurs before the $V_c$ jet, and during these periods
of time a proper distance cannot be measured. Then, the jet appears near the linear
polarization patch, allowing one to measure the distance between them. Finally,
some cases show that the linear polarization patch vanishes before the $V_c$ jet,
providing a situation where the distances between them cannot be measured again.
 
Figure \ref{fig3} provides the histogram of these distances with a mean
value of 0.\arcsec57.  These distances are measured in the non-restored data because
linear polarization patches are not easily detected in the restored data (due to the
increased noise in them).  Thus, to compare this with the results of the previous
section, one should use the values from the bottom row in Fig.  \ref{fig2}.  In
particular, the mean distance of $V_c$ jets to neutral lines in the
non-restored data is 0.\arcsec41. This value is lower than the
mean distance between the center of the $V_c$ jets and the center of the linear
polarization patch. This
indicates that the jets have, on average, a neutral line closer to them than
the one corresponding to the associated linear polarization patch. This
tendency has been confirmed by visual inspection in 80\% of the cases analyzed.
The remaining 20\% are different in the sense that the closest neutral line to
the $V_c$ signal is the one of the associated linear polarization patch. Below
we discuss different examples of these two types.

\begin{figure*}
\centering
\includegraphics[width=18cm]{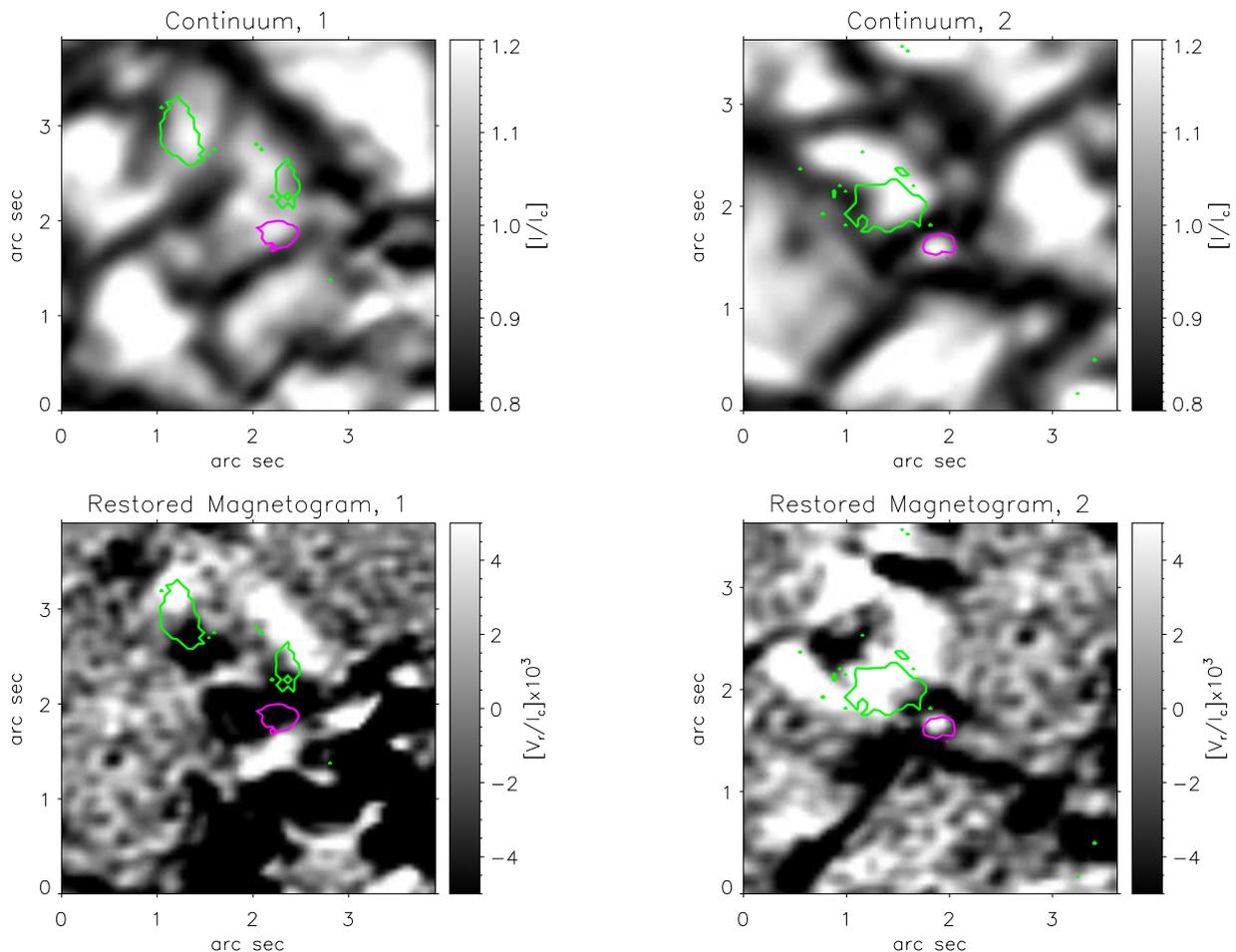}
\caption{Two examples of the most common configuration
found (80 \% of the associated events) where the linear polarization patch (green)
is closer to a neutral line different from the one that is nearest to the $V_c$ jet (purple).
The top row displays a continuum map in which the linear polarization
signals always appear at the edges of granules. The bottom row shows the restored magnetograms.
This configuration exemplifies the conclusions from Figs. \ref{fig2} and \ref{fig3}.}
\label{fig4}
\end{figure*}

\begin{figure*}
\centering
\includegraphics[width=18cm]{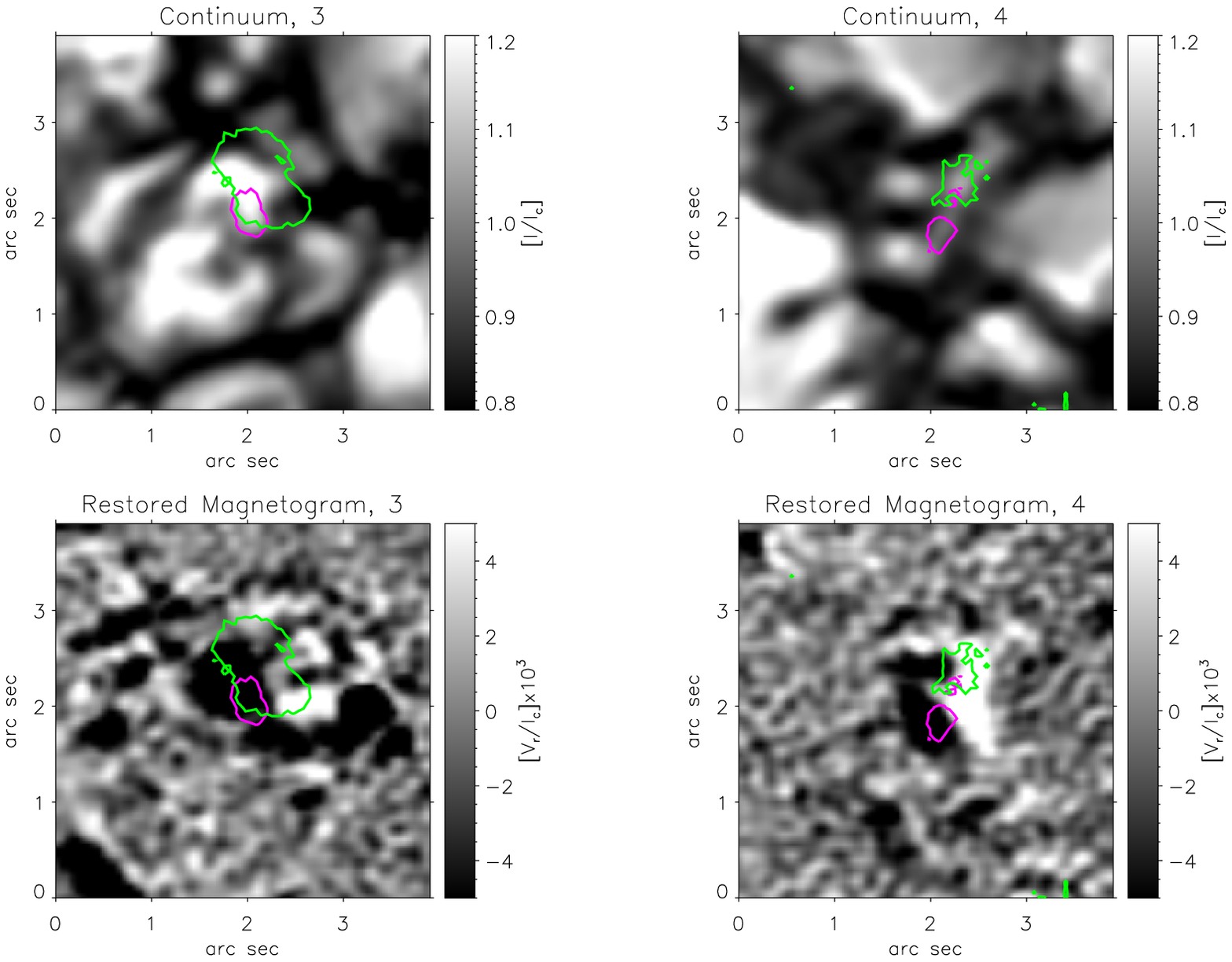}
\caption{Two examples of the less common configuration (only 20\% of the associated
cases) where the linear polarization patch (green) and the $V_c$ jet (purple)
are related to the same neutral line configuration. The images displayed correspond to the
same quantities as given in Fig. \ref{fig4}.}
\label{fig5}
\end{figure*}

In Fig. \ref{fig4}, two events of the first type are shown. In them, a newly 
emerged loop, identified by the transverse fields (green contours) and the two footpoints 
of opposite polarity in the underlying (restored) magnetogram, generate a $V_c$ jet 
(purple contour). The jet is closer to a neutral line different from the one associated to the
loop's footpoints. This other neutral line is created by the interaction of 
one of the footpoints with a pre-existing field region of opposite polarity.
In the first example in this figure, two loops are seen. The jet is closer to
the smaller one in the middle of the image. The top footpoint of this loop is of positive polarity
and the bottom one of negative polarity.

This last footpoint encounters
a small positive polarity field region and the $V_c$ jet is seen on this side
of the loop, on top of the negative polarity footpoint. It seems natural to associate 
the jet with this closer neutral line and not with the one overlying the loop itself.
The second example corresponds to a
larger transverse patch where the negative polarity of the loop encounters
a tiny round positive polarity element of 0.\arcsec2 ~diameter that seems to be 
responsible for the $V_c$ jet in this case. Note the many other lines in both pictures
without any linear polarization (which consequently are not newly emerged quiet-Sun loops) or $V_c$ signals. 

Figure \ref{fig5} shows two examples of newly emerged loops that display $V_c$ jets whose
position is closer to the neutral line created by the 
loop itself (representing 20\% of the cases). The first example shows a large transverse field 
patch that intersects most of the $V_c$ jet area. The loop is asymmetric and the negative polarity occupies
a larger area than the positive one. Even if other neutral lines are seen nearby (to the
right of the linear polarization patch), the jet is closer to the neutral line that is
seen inside the green contour. A more obvious case is displayed in the second
example of this figure, where an isolated loop is shown with both the green
and purple contours intersecting the neutral line that is created by the opposite
polarity footpoints of the loop. No other observed neutral line can be argued
to be responsible for the jet signal in this case. 

It is also of interest how systematically the transverse
field regions are located at the borders of granules \citep[]{Lites1996, Lites2008, 
Danilovic2010_2}, as shown by the continuum frames in Figures \ref{fig4} and \ref{fig5}.

\subsection{Description of a complex case}

While the majority of the jets observed in these two time
series correspond to an association of one linear polarization 
patch and a single $V_c$ feature, a few cases have been observed
that show multiple high-speed flows triggered by linear polarization
patches. We describe here the most complex of these associations observed with
IMaX/SUNRISE, including its evolution, and how the jets respond to the
varying granulation pattern.

A clarification is in order beforehand. The circular polarization
signals in the continuum, $V_c$, are a signed quantity. So far we
have not discussed the signs of the observed jets. The example explained here displays $V_c$ of both signs.
However, in the case of the IMaX observation, it is hard to make a sensible
use of the observed signs of the $V_c$ in these jets. Because the continuum point used
in this instrument is midway between the Fe~{\sc i} 5250.2 \AA~and the
Fe~{\sc i} 5250.6 \AA~lines, a blueshift or redshift can give rise to oppositely
signed signals at $\lambda_c$. Besides, if one compares jets
with the same flow directions but opposite magnetic orientations
of the line-of-sight component, the signs of the
corresponding $V_c$ signals are reversed as well. Thus, if we observe two $V_c$ jets
next to each other with different signs, it is impossible to distinguish between
opposite flow directions and/or opposite field line orientations. Data with
better spectral coverage are needed to achieve this. One could assume that the sign
of the magnetic or velocity field in the jet atmosphere is the same as that 
inferred from the four points inside the spectral line. But this assumption
is not fully justified before the nature of quiet-Sun jets is better 
understood.

\begin{figure*}
\centering
\includegraphics[width=23.5cm,angle=90]{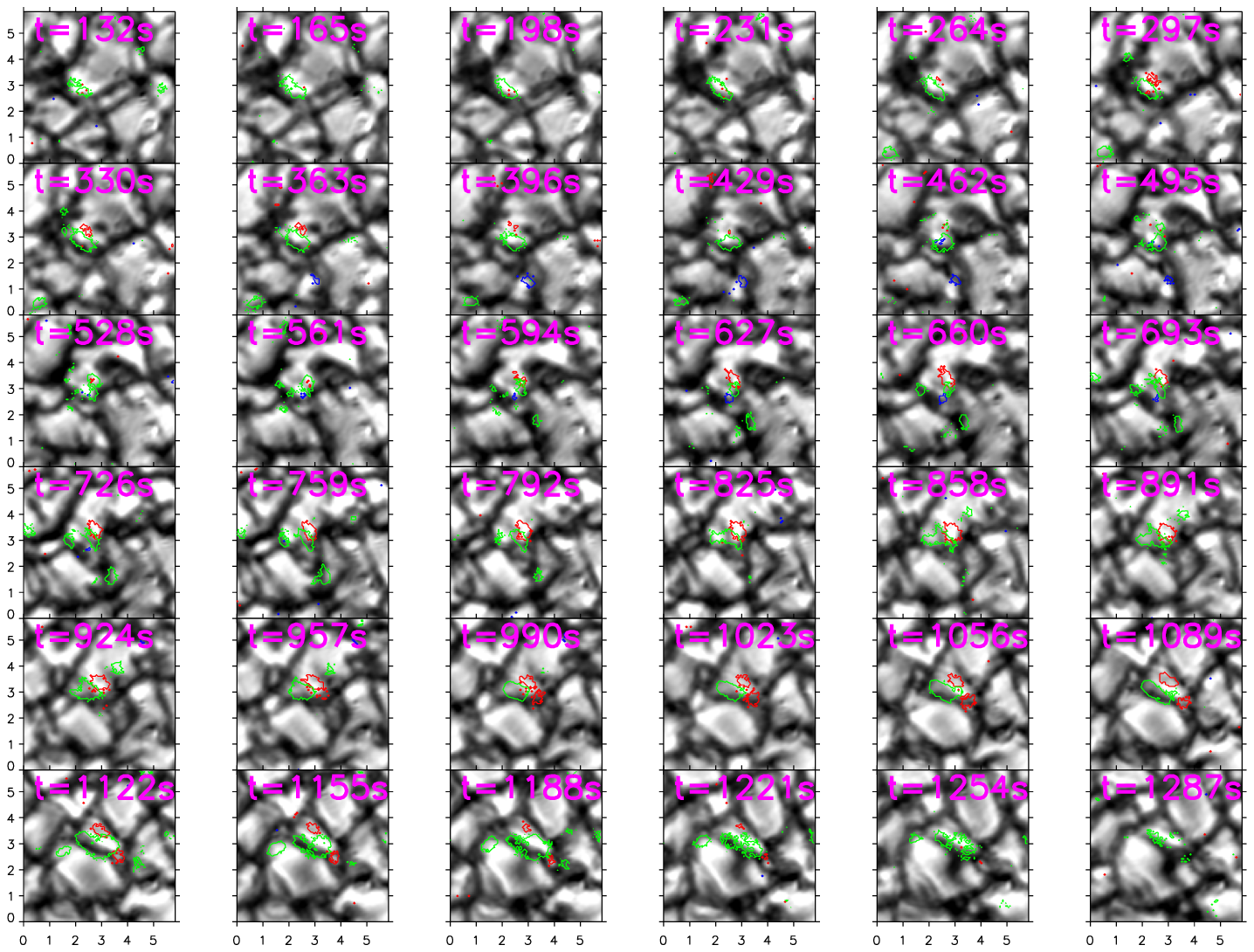}
\caption{Evolution of the most complex $V_c$ jet observed 
with IMaX/SUNRISE. The images correspond to the continuum
intensity. Linear polarization is represented by green contours over 
the granular pattern. Blue
and red contours correspond to positive and negative $V_c$ signals. Time
(given in each frame) runs from left to right, starting at the top (as seen by
turning the page on its side). Units given with the maps are in arcsec.
} \label{fig6}
\end{figure*}

\begin{figure*}
\centering
\includegraphics[width=23.5cm,angle=90]{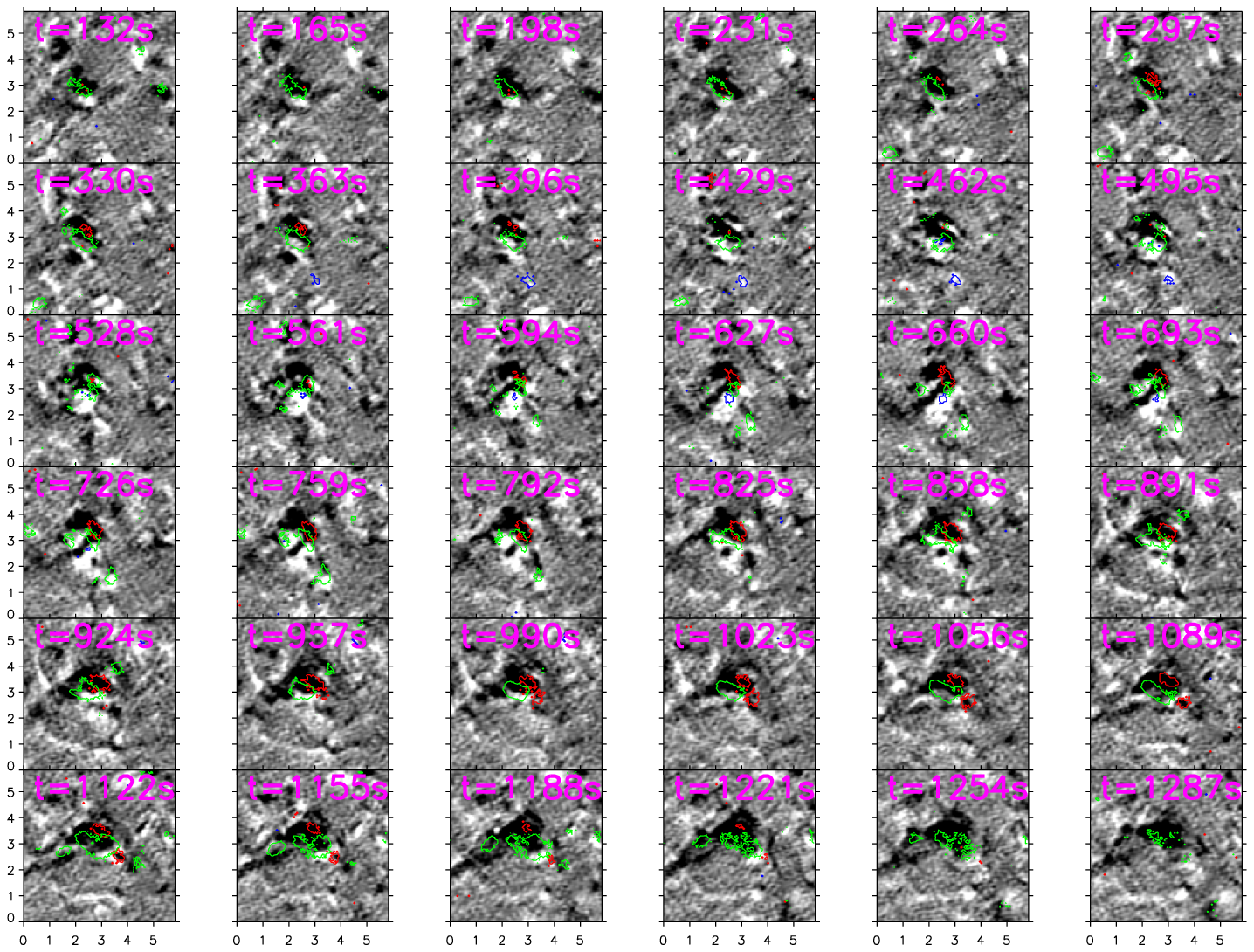}
\caption{Same as Fig. \ref{fig6}, but with the reconstructed magnetograms
as background images.} \label{fig7}
\end{figure*}

Figure \ref{fig6} covers almost 20 minutes of a $6\times6$ arc sec$^2$ area from the first time series. In this area,
the most complex jet event observed by IMaX/SUNRISE was detected.  The
background of each snapshot corresponds to the continuum intensity and allows
examining the evolution of the granular pattern during the event.  As before,
green contours designate the linear polarization signals, but now the $V_c$
contours indicate the sign of the events in blue (positive $V_c$) and red (negative $V_c$).
The event develops mostly negative $V_c$ jets but also creates positive $V_c$ signals for a
short period of time. As explained before, this can only be
taken as an indication that either oppositely directed strong flows or opposite
polarity fields participate in the event. Time is given in seconds from
the beginning of the full observed data set. The corresponding
reconstructed magnetograms are shown in Fig. \ref{fig7}.

A large patch of linear polarization is first seen at $t=132s$ located
at the edge of a granule and progressively increases in size. Since the 
first snapshot was taken a few scattered pixels near that patch show negative signal at $\lambda_c$ (see 
the red pixels close to the patch). At $t=330s$, the event reaches a temporary maximum 
with a clear jet-transverse field association. After this time, the transverse field
patch slightly fades, but never disappears completely. Interestingly, at
$t=462s$, a positive polarity jet starts to develop and attains a maximum at
$t=627s$. Positive/negative associations of jets with clear upflow/downflow components
were observed in \cite{MartinezPillet2011a} using Hinode/SP data.
While this positive $V_c$ component disappears two snapshots later, the negative (red)
component of the jet and the transverse field patch again increase.
The strongest development of the event occurs in the interval $t=[858,1122]s$. 
In this interval, around $t=990s$, the negative $V_c$ jet has broken into two adjacent patches 
that move in opposite directions as the horizontal field patch increases 
in size. The two periods with increasing linear polarization signals coincide
with a similar increase of the associated granule. Periods of decreasing linear 
polarization signal ($t=[429,627]s$) coincide with granular break-up. This reinforces
the relation of these events with flux emergence episodes connected with granular upflows. 
From a magnetic point of view (see Fig. \ref{fig7}), this case is different from the previous
ones studied in the sense that, now, a multipolar structure seems to be emerging.
At $t=495s$ or at $t=693s$, two patches with opposite polarity footpoints are seen. The system
only resolves into a more or less bipolar structure at the very end, in the second
phase of increasing linear polarization signals ($t=1188s$). In the present case,
no pre-existing fields seem to be playing a major role (although they cannot
be completely ruled out), and one would favor jets triggered by interactions
of the intrinsically tangled (multipolar) topology of the magnetic structure.

Figure \ref{fig6} also shows the onset of an isolated blue jet at the edge of a granule 
 near the bottom of the $t=363s$ snapshot. This strong flow is generated
with no clear connection to any linear polarization patch and constitutes an
isolated jet. It stays visible until $t=495s$ (see Fig. \ref{fig6}).
Interestingly, a loop is seen at a later time $t=528s$ (see Fig. \ref{fig7}).
We therefore assume that here a transverse field emerged at
the time at which the jet was detected, but it did not produce strong enough Stokes
$Q$ and/or $U$ signals to be detected by IMaX/SUNRISE.

\section{Discussion}

The evidence presented in this paper supports the original idea presented in
\citet{Borrero2010,Borrero2012} and \citet{MartinezPillet2011a} for the origin
of the quiet-Sun jets detected by IMaX/SUNRISE. Granular upflows are connected with
surfacing small-scale flux loops. The first evidence of these
loops was obtained from a linear polarization patch that preceded any signature of a strong flow
most of the time. These flux loops often have a bipolar
configuration but, sometimes, more complex configurations are encountered. The
neutral lines created by the loop's footpoints are naturally observed below
them at the spatial resolution of the observations. Most jets are systematically
located at larger distances from the transverse field regions than from
nearby neutral lines formed between the emerging fields and
pre-existing ones.  In a small number of cases no transverse
fields are seen before the jet, which might be explained by 
a lack of sensitivity to the linear polarization signals. In approximately 20\%
of the cases we were unable to identify  a nearby neutral line with a pre-existing field. 
We speculate that in these events we missed a weak pre-existing field due to insufficient
sensitivity to the circular polarization signals. Thus, the sequence of
flux loop emergence and interaction with pre-existing fields might be valid to
explain more or less all quiet-Sun jets. 

The physical mechanism that is most naturally invoked to explain high-speed flows, the
$V_c$ jets, is magnetic reconnection. The recent work of
\citet{Borrero2013} has shown from full radiative-transfer Stokes inversions that
magnetic fields with different inclinations along the line-of-sight and temperature
increases are needed to fit the profiles.  This strongly favors reconnection as
the underlying mechanism. 

The most complex case observed, which showed a
multipolar configuration, might be better explained as interactions among the
different polarity footpoints of the structure itself that also lead to
reconnection processes.  There is a clear tendency for both the transverse
field patches and the $V_c$ jets to appear near the edges of granules
\citep[the former result dates back to][]{Lites1996}. Their
evolution follows the granular expansion. All in all, it would seem
that the loops interact and reconnect with pre-existing fields after being
pushed by the granules to the edges, where they encounter the fields in the
nearby intergranular regions. There reconnection events are finally triggered.
$V_c$ jets of opposite sings (in the observed circular
polarization at $\lambda_c$) naturally arise from the coexistence of different
field orientations and oppositely directed flows as a result of the
reconnection. 

We note that because the emerging fields seem to be passively advected by the
granulation, the field strengths should be around or below equipartition
with the convective flows.
The results presented here and in the previously cited works indicate
that these weak fields are nonetheless capable of generating the observed 
high-speed flows when reconnecting.
Note, however, that the pre-existing field also involved in the reconnection
may well be strong \citep[see][]{Lagg2010}.
While further studies are needed to consolidate magnetic 
reconnection as the physical mechanism responsible for them, other
possibilities should not be excluded. The intrinsically larger number of
redshifted events found by \citet{MartinezPillet2011a} indicates that 
strong downflows may be initiated by some other physical mechanism, such as
convective collapse \citep{Parker1978, Grossmann1998, Nagata2008, Shimizu2008, 
Fischer2009, Danilovic2010_1}. The 
association between upflows and downflows encountered  
by \citet{MartinezPillet2011a} could be reminiscent of some form 
of siphon-flow process \citep{Rueedi1992,Montesinos1993}. Finally, it is known that
granular evolution itself leads to supersonic (horizontal) flows 
\citep{Cattaneo1989,Solanki1996,Bellot2009} that could play a role
in generating the observed $V_c$ events.

\section{Conclusions}

We quantitatively analyzed the relation between quiet-Sun
jets described by \citet{Borrero2010} and the transverse field patches
observed with IMaX/SUNRISE. In particular, the timing relation between
them and the distances of these features to nearby 
neutral lines observed in magnetograms
were quantified. The results can be summarized as follows:

In 77\% of the cases for which a clear association between transverse
fields and $V_c$ jets could be established, we found that the transverse
fields occur before the jets, on average by about 84 seconds, and
always by less than 300 seconds, a typical granulation timescale.

$V_c$ jets have a mean distance to nearby neutral lines of 0.\arcsec19.
This is significantly shorter than the distance of randomly distributed points 
in IMaX/SUNRISE magnetograms from neutral lines 
(0.\arcsec5). This reinforces the association of these processes with 
reconnection events. However, 0.\arcsec19 is larger than the spatial
resolution of the data, so that the jets are associated with neutral lines, but not
necessarily overlap them.

The transverse field patches have a typical distance of 0.\arcsec16
to a neutral line, which is close to the spatial resolution of the data. Because they occur more
often over granular regions, these patches are more easily interpreted
in terms of magnetic loop emergence \citep{Danilovic2010_2}.

The distance between $V_c$ jets and the transverse patches is typically 
 0.\arcsec57  (as measured in the non-reconstructed data), larger than the 
typical distance between jets and the nearest neutral lines. 

Visual inspection showed that 80\% of the $V_c$ jets are closer to neutral lines that are generated between
one of the loop's footpoints and a pre-existing field of opposite polarity
than to the neutral line between the loop footpoints. The remaining
20\% of the $V_c$ jets can only be associated to the neutral
line of the recently emerged loop. We cannot rule out that we missed pre-existing
fields in these cases because of the limited sensitivity of the observations.

Complex cases with multipolar configurations and repeated $V_c$ jet
events can also occur.

Both the transverse field regions and the $V_c$ jets evolve following
the underlying granular pattern, in particular, they occur near
the edges of the granules most of the time.

All these results support the suggestion that quiet-Sun jets are a 
result of the interaction of a newly emerged granular loop with already 
existing fields, which triggers magnetic reconnection. One observable signature of
this reconnection are the $V_c$ jets. It is clear, however, that to fully characterize
these processes, analyses of data with better spectral coverage that include
several magnetically sensitive lines are needed. In addition, a study of the signatures
of these jets at different limb distances would be very useful.
Work is in progress to perform such studies using Hinode and SUNRISE data,
to obtain a better understanding of the physical process. In addition,
an MHD simulation of emerging magnetic flux and its interaction with a pre-existing
field would provide valuable physical insight. 

￼\begin{acknowledgement}
We thank Eric Priest for interesting discussions on the
possible nature of these quiet-Sun jets.
This work has been partially funded by the Spanish
MINECO through Project No. AYA200AYA2011-29833-C06 and
by the WCU grant No.R31-10016 funded by the Korean
Ministry of Education, Science and Technology. 
￼\end{acknowledgement}

\bibliographystyle{aa} 
\bibliography{cqnbib.bib} 

\end{document}